%
%



\newcount\chno \chno=0
\newcount\equno \equno=0
\newcount\refno \refno=0

\font\chhdsize=cmbx12 at 14.4pt

\def\startbib{\def\biblio{\bigskip\medskip
  \noindent{\bf REFERENCES.}\bgroup\parindent=2em}}
\def\endbib{\edef\biblio{\biblio\egroup}}
\def\reflbl#1#2{\global\advance\refno by 1
  \edef#1{\number\refno}
    \global\edef\biblio{\biblio\smallskip\item{[\number\refno]}#2\par}}

\def\eqlbl#1{\global\advance\equno by 1
  \global\edef#1{{\number\chno.\number\equno}}
  (\number\chno.\number\equno)}


\def\qed{\hfil\hbox to 0pt{}\ \hbox to 2em{\hss}\ 
         \hbox to 0pt{}\hskip-2em plus 1fill
         \vrule height6pt depth1pt width7pt\par\medskip}
\def\eqed{\hfil\hbox to 0pt{}\ \hbox to 2em{\hss}\ 
          \hbox to 0pt{}\hskip-2em plus 1fill
          \vbox{\hrule height .25pt depth 0pt width 7pt
            \hbox{\vrule height 6.5pt depth 0pt width .25pt
              \hskip 6.5pt\vrule height 6.5pt depth 0pt width .25pt}
            \hrule height .25pt depth 0pt width 7pt}\par\medskip}


\def\msimp#1#2{#1%
  \hbox to 0pt{\hskip 0pt minus 3fill
    \phantom{$#1$}\hbox to 0pt{\hss$#2$\hss}\phantom{$#1$}%
    \hskip 0pt minus 1fill}}


\def\rhs{{right-hand side}}

\def\dif#1#2{{\partial_{#2} #1}}

\def\bj{{\bf j}}
\def\bn{{\bf n}}
\def\bp{{\bf p}}

\def\bs{{\bf s}}

\def\bv{{\bf v}}
\def\bw{{\bf w}}
\def\bx{{\bf x}}
\def\by{{\bf y}}

\def\bom{\bw}

\def\bB{{\bf B}}
\def\bE{{\bf E}}
\def\bJ{{\bf J}}

\def\bP{{\bf P}}

\def\bX{{\bf X}}
\def\bY{{\bf Y}}
\def\bZ{{\bf Z}}

\def\cI{{\cal I}}
\def\cM{{\cal M}}


\font\tendouble=msbm10 \font\sevendouble=msbm7  
\font\fivedouble=msbm5

\newfam\dbfam
\textfont\dbfam=\tendouble \scriptfont\dbfam=\sevendouble
\scriptscriptfont\dbfam=\fivedouble

\mathchardef\dbB="7042 
\mathchardef\dbC="7043 
\mathchardef\dbD="7044 
\mathchardef\dbE="7045 
\mathchardef\dbF="7046 
\mathchardef\dbG="7047 
\mathchardef\dbH="7048 
\mathchardef\dbI="7049 
\mathchardef\dbJ="704A 
\mathchardef\dbK="704B 
\mathchardef\dbL="704C 
\mathchardef\dbM="704D 
\mathchardef\dbN="704E \def\NN{{\fam=\dbfam\dbN}}
\mathchardef\dbO="704F 
\mathchardef\dbP="7050 
\mathchardef\dbQ="7051 
\mathchardef\dbR="7052 \def\RR{{\fam=\dbfam\dbR}}
\mathchardef\dbS="7053 \def\SS{{\fam=\dbfam\dbS}}
\mathchardef\dbT="7054 
\mathchardef\dbU="7055 
\mathchardef\dbV="7056 
\mathchardef\dbW="7057 
\mathchardef\dbX="7058 
\mathchardef\dbY="7059 
\mathchardef\dbZ="705A 


\startbib

\reflbl\abraham{
	M. Abraham, 
		{\it Prinzipien der Dynamik des Elektrons},
	Ann. Phys. {\bf 10}, pp. 105--179 (1903).}

\reflbl\abrahamBOOK{
	M. Abraham, 
		{\it Theorie der Elektrizit\"at, II}, 
	Teubner, Leipzig (1905,1923).}

\reflbl\appelkiessling{
	W. Appel and M.K.-H. Kiessling,
		{\it in preparation} (1999).}

\reflbl\barutBOOKa{
	A.O. Barut, 
		{\it Electrodynamics and classical theory of fields
			and particles}, 
	Dover, New York (1964).}

\reflbl\barutBOOKb{
	A.O. Barut (Ed.),
		{\it Foundations of radiation theory and Quantum 
		Electrodynamics},	
	Plenum, New York (1980).}

\reflbl\bauerduerr{
	G. Bauer and D. D\"urr, 
		{\it The Maxwell-Lorentz system of a rigid charge
		distribution},
	Preprint, LMU M\"unchen (1999).}
 
\reflbl\dirac{
	P.A.M. Dirac,	
		{\it Classical theory of radiating electrons},
	Proc. Roy. Soc. A {\bf 167}, p. 148 (1938).}

\reflbl\einstein{
	A. Einstein,	
	{\it Zur Elektrodynamik bewegter K\"orper},
	Ann. Phys. {\bf 17}, pp. 891ff. (1905).}

\reflbl\dysonA{
	F. Dyson,	
		{\it The radiation theories of Tomonaga, Schwinger,
		and Feyman,}
	Phys. Rev. {\bf 75}, pp. 486--502 (1949); {\it The S matrix in
	Quantum Electrodynamics}, ibid. pp. 1736--1755.}

\reflbl\dysonB{
	F. Dyson,	
		{\it Divergence of perturbation theory in Quantum
		Electrodynamics},
	Phys. Rev. {\bf 85}, pp. 631--632 (1952).}

\reflbl\fermi{
	E. Fermi, 	
		{\it \"Uber einen Widerspruch der elektrodynamischen und 
		der relativistischen Theorie der elektromagnetischen Masse},
	Phys. Zeitschr. {\bf 23}, pp. 340--344 (1922).}

\reflbl\jacksonBOOK{
	J.D. Jackson, 
		{\it Classical Electrodynamics}, 3rd ed.,
	Wiley, New York (1999).}

\reflbl\jauchrohrlichBOOK{
	J.M. Jauch  	
	and 
	F. Rohrlich,	%
		 {\it The theory of photons and electrons},
	Springer, NY (1976).}

\reflbl\komechspohnkunze{
	A. Komech, H. Spohn, and M. Kunze, 
		{\it Long-time asymptotics for a classical particle 
		interacting with a scalar wave field}, 
	Commun. PDE {\bf 22}, 307--335 (1997).}

\reflbl\komechkunzespohn{
	A. Komech, M. Kunze, and  H. Spohn, 
		{\it Effective dynamics of a mechanical particle coupled
		to a wave field},
	to appear in Commun. Math. Phys. {\bf } (1999).}

\reflbl\komechspohn{
	A. Komech and  H. Spohn, 
		{\it Long-time asymptotics for the coupled
		Maxwell-Lorentz Equations}, 
	to appear in J. Diff. Eq. {\bf } (1999).}

\reflbl\kunzespohnA{
	M. Kunze and H. Spohn,
		{\it Adiabatic limit of the Maxwell-Lorentz equations},
	Preprint, TU M\"unchen (1998).}

\reflbl\kunzespohnB{
	M. Kunze and H. Spohn,
		{\it Radiation reaction and center manifolds},
	Preprint, TU M\"unchen (1999).}

\reflbl\lienard{
	A. Li\'enard, 
		{\it Champ \'electrique et magn\'etique produit
		par une charge concentr\'ee en un point et anim\'ee
		d'un mouvement quelconque,}
	L'\'eclairage \'Electrique {\bf 16} p.5;
	ibid. p. 53; ibid. p. 106 (1898).}

\reflbl\lorentzforce{
	H.A. Lorentz, 
		{\it La th\'eorie \'electromagnetique de Maxwell et son 
			application aux corps moevemants},
	Arch. N\'eerl. Sci. Exactes Nat. {\bf 25}, pp. 363--552 (1892).}

\reflbl\lorentzBOOKa{
	H.A. Lorentz, 
		{\it Versuch einer Theorie der elektrischen und
		optischen Erscheinungen in bewegten K\"orpern},
	Teubner, Leipzig (1909) (orig. Leyden (1895).)}
	
\reflbl\lorentzENCYCLOP{
	H.A. Lorentz, 
		{\it Weiterbildung der Maxwell'schen Theorie:
			Elektronentheorie.},
	Enzyklopaedie d. Mathematischen Wissenschaften ${\bf V}_2$,
	pp. 145--280 (1904).}

\reflbl\lorentzACAD{
	H.A. Lorentz, 
		{\it Electromagnetic phenomena in a system moving
		with any velocity less than that of light},
	Proc. Acad. Wet. Amsterdam, {\bf 6}, pp. 809ff. (1904).}

\reflbl\lorentzBOOKb{
	H.A. Lorentz, 
		{\it The Theory of electrons and its applications 
			to the phenomena of light and radiant heat}, 
	2nd ed., 1915; reprinted by Dover, New York (1952).}

\reflbl\minkowski{
	H. Minkowski, 	
		{\it Die Grundgleichungen f\"ur elektromagnetische
		Vorg\"ange in bewegten K\"orpern},
	G\"ottinger Nachr. {\bf} pp. 53 ff. (1908).}  

\reflbl\nodvik{
	J.S. Nodvik, 	
		{\it A covariant formulation of Classical Electrodynamics
			for charges of finite extension},
	Ann. Phys. (N.Y.) {\bf 28}, pp. 225--319 (1964).}

\reflbl\peierls{
	R. Peierls, 	
	{\it More surprises in theoretical physics},
	Princeton Univ. (1991).}

\reflbl\poincare{
	H. Poincar\'e,	
		{\it Sur la dynamique de l'\'electron},
	Comptes Rendus {\bf 140}, pp. 1504--1508 (1905);
	Rendiconti del Circolo Matematico di Palermo {\bf 21}, 
	pp. 129--176 (1906).}

\reflbl\rohrlichBOOK{
	F. Rohrlich,	
		{\it Classical charged particles}, 
	Addison Wesley, Redwood City, CA (1990).}

\reflbl\rohrlichAJPold{
	F. Rohrlich,	
		{\it Self-energy and stability of the classical electron},
	Am. J. Phys. {\bf 28}, pp. 639--643 (1960).}

\reflbl\rohrlichAJPnew{
	F. Rohrlich,	
		{\it The dynamics of a charged sphere and the electron},
	Am. J. Phys. {\bf 65}, pp. 1051--1056 (1997).}

\reflbl\schwinger{
	J. Schwinger, 	
		{\it Electromagnetic mass revisited},
	Found. Phys. {\bf 13}, pp. 373--383 (1983).}

\reflbl\sommerfeldBOOK{
	A. Sommerfeld, 
	{\it Electrodynamics}, 
		Academic Press, New York (1952).}

\reflbl\spohn{
	H. Spohn, 	
		{\it Runaway charged particles and center manifold theory},
	Preprint, TU Munich (1998).}

\reflbl\thomas{
	L.H. Thomas,	
		{\it The motion of the spinning electron},
	Nature {\bf 117}, p. 514 (1926);
		{\it On the kinematics of an electron with an axis},
	Phil. Mag. {\bf 3}, pp. 1--22 (1927).}
 
\reflbl\jjthomsonA{
	J.J. Thomson, 
		{\it On the electric and magnetic effects produced by
		the motion of electrified bodies}, 
	Phil. Mag. {\bf 11}, pp. 227 ff. (1881).}
 
\reflbl\jjthomson{
	J.J. Thomson, 
	{\it Cathode rays}, Phil. Mag. {\bf 44}, pp. 294--316 (1897).}

\reflbl\tomonagaBOOK{
	S. Tomonaga,
		{\it The story of spin},
	Univ. Chicago Press (1997).}

\reflbl\uhlenbeckgoudsmit{
	G. E. Uhlenbeck and S. A. Goudsmit, 
		{\it Spinning electrons and the structure of spectra},
	Nature {\bf 117}, pp. 264--265 (1926).}

\reflbl\wiechert{
	E. Wiechert, 
		{\it Elektrodynamische Elementargesetze}, 
	Arch. N\'eerl. Sci. Exactes Nat. {\bf 5}, pp. 549 (1900).}

\reflbl\yaghjian{
	A. D. Yaghjian,
		{\it Relativistic dynamics of a charged sphere},
	Lect. Notes Phys. {\bf m11}, Springer, Berlin (1992).} 

\endbib

\magnification=\magstep1


\centerline{\chhdsize CLASSICAL ELECTRON THEORY}
\smallskip
\centerline{\chhdsize AND CONSERVATION LAWS }

\bigskip
\centerline{MICHAEL K.-H. KIESSLING}
\medskip
\centerline{Department of Mathematics}
\centerline{Rutgers University}
\centerline{110 Frelinghuysen Rd.}
\centerline{Piscataway, NJ 08854, USA}

\bigskip
\noindent
{\bf ABSTRACT:}  
	{It is shown that the traditional conservation laws for 
	 total charge, energy, linear and angular momentum, hold 
	 jointly in classical electron theory if and only if classical
	 electron spin is included as dynamical degree of freedom.}

\medskip

\centerline{To appear in: {\bf Physics Letters A}}


\bigskip
\noindent
{\bf I. INTRODUCTION}\chno=1
\medskip

`Classical electron theory' aims at a consistent description of
 the dynamics of electromagnetic fields and charged particles. 
While already equipped with some heuristic rules to the effect 
 that charged particles occur in stable, `quantized' atomic units, 
 classical electron theory is dynamically a pre-quantum theory, 
 built on Maxwell's electromagnetism, Newton's or 
 Einstein's classical mechanics, and the Lorentz force law [\lorentzforce].
It originated around the time of J.J. Thomson's discovery of the
 corpuscular nature of cathode rays [\jjthomson] with the work of 
 H. A. Lorentz [\lorentzBOOKa], just before the advent of relativity 
 [\lorentzACAD,\einstein,\poincare,\minkowski], 
 and is most prominently associated with the names of Lorentz 
 [\lorentzBOOKa,\lorentzENCYCLOP] and Abraham [\abraham] 
 (see also [\abrahamBOOK,\lorentzBOOKb,\sommerfeldBOOK]). 
Since the simplest classical mathematical structure for an atomic charge,
 the point charge, produces infinite Lorentz self-forces and
 electromagnetic self-energies associated with the singularities of 
 the Li\'enard-Wiechert electromagnetic field [\lienard,\wiechert] 
 of a moving point charge, Abraham and Lorentz 
 were led to consider models with extended microscopic 
 charges that experience a volume-averaged Lorentz force.
While problems of how to bring classical electron theory in line 
 with special relativity still persisted [\poincare,\lorentzBOOKb,\fermi], 
 quantum mechanics was invented, and classical electron 
 theory dropped from the list of contenders 
 for a fundamental theory of electromagnetic matter and fields. 
Nowadays classical electron theory has degenerated into 
 a subject of mere historical value --- so it would seem, 
 especially in view of the impressive range of 
 electromagnetic phenomena covered with spectacular precision 
 by Quantum Electrodynamics (QED) [\jauchrohrlichBOOK,\barutBOOKb]. 
However, classical electron theory has continually been revisited
 by physicists, an incomplete list being
 [\fermi,\dirac,\rohrlichAJPold,\nodvik,\rohrlichAJPnew,\schwinger,\yaghjian,\jacksonBOOK].
One reason, apparently, is that the perturbative series defining QED, 
 while renormalizable in	each order [\dysonA], is most likely 
 to be merely asymptotic in character rather than convergent 
 [\dysonB], such that precision results are achieved only when 
 computations are terminated after a few iterations, with no 
 rules as to what is meant by `few'. 
As R. Peierls put it, [\peierls]: ``because the use, 
 as basic principle, of a semiconvergent series is unsatisfactory, 
 there is still interest in theories that avoid singularities.''
\smallskip
{\hrule\medskip
\noindent
$\msimp{\copyright}{c}$ (1999) The author. 
Reproduction for non-commercial purposes of this article, 
in its entirety, by any means is permitted. (The copyright 
will be transferred to Elsevier for publication in Phys. Lett. A.)}

Recently, considerable advances have  been made in controlling the type
 of dynamical equations that govern classical electron theory.
A mechanical particle coupled to a scalar wave field has been treated in 
 [\komechspohnkunze,\komechkunzespohn,\kunzespohnB]. 
The equations of a classical electron coupled to the Maxwell fields 
 are considered in [\bauerduerr,\komechspohn,\kunzespohnA,\spohn]. 
All these papers deal with a semi-relativistic theory. 
This means that, while the wave field satisfies relativistic equations 
 and the `material' particle momentum is given by Einstein's rather 
 than Newton's expression, with Abraham [\abrahamBOOK] one assumes that 
 the particle rigidly retains its shape. 
A fully relativistic model, first devised in a monumental work
 by Nodvik [\nodvik] and most recently completed in
 [\appelkiessling], is considerably more involved but has
 begun to yield to a mathematical onslaught as well [\appelkiessling]. 
As a result, classical electron theory is about to become
 established as the first mathematically well posed, fully 
 relativistic theory of electromagnetism that consistently describes the 
 dynamics of discrete charges and the continuum electromagnetic  field.

In view of the continued interest in classical electron theory 
 it seems worthwhile to draw attention to a 
 small observation regarding conservation laws which, to the 
 author's knowledge, has not been made before, and which seems 
 to be sufficiently interesting in its own right to warrant 
 publication in this separate note. 

To be specific, 
 [\bauerduerr,\komechspohn,\kunzespohnA,\spohn,\komechspohnkunze,\komechkunzespohn,\kunzespohnB]
 and most earlier works on classical electron theory only take 
 translational degrees of freedom of the particles into account. 
Already Abraham [\abrahamBOOK] and Lorentz [\lorentzACAD] insisted 
 on the possibility of additional degrees of freedom  of the extended
 charged particles associated with particle spin (not to be confused 
 with the ``rotation of the electrons'' in  Lorentz' theory of  
 the Zeeman effect [\lorentzBOOKb], which refers to circular
 motion of the electron's center of charge inside a Thomson atom),
 though it seems that only Abraham wrote down corresponding 
 dynamical equations.
However, neither of these authors pursued such a 
 spinning particle motion any further. 
As a consequence, it seems to have gone completely unnoticed
 that omitting particle spin generally leads to a violation of 
 the law of conservation of total classical angular momentum!
The discrepancy term has the form of an internal torque on the particles. 
This strongly suggests to add classical spin to the 
 degrees of freedom of the charge distribution.

In  this note we will show that, if classical particle spin is 
 included as degree of freedom in semi-relativistic classical 
 electron theory, with Abraham's spherical charge distribution, 
 then all classical conservation laws are satisfied. 
We will also demonstrate that the arbitrary setting to zero 
 of the internal angular velocities of the particles is incompatible
 with the classical law of angular momentum conservation. 
Interestingly, though, the classical expressions for total 
 charge, energy, and linear momentum are conserved 
 during the motion even if classical particle spin is omitted.

In a fully relativistic formulation [\nodvik,\appelkiessling] the 
 kinematical effect of Thomas precession [\thomas] enforces
 additional self rotation of the particle. 
However, it is the law of angular momentum conservation which
 {\it compels us} to introduce a `classical particle spin' 
 already at the level of a semi-relativistic formulation of 
 classical electron theory, i.e. independendly of Thomas precession.  

The rest of the paper is organized as follows. 
We first  present the dynamical equations (section II), next we prove 
 that the traditional conservation laws are satisfied (section III), 
 then we show (section IV) that angular momentum is not conserved if 
 spin is omitted.
In section V, we conclude with a brief historical musing.


\bigskip
\noindent
{\bf II. THE EQUATIONS OF SEMI-RELATIVISTIC ELECTRON THEORY }
\chno=2\equno=0
\medskip
 
We denote by $\bE (\bx , t)\in\RR^3$ the electric field, 
 and by $\bB(\bx ,t)\in\RR^3$ the magnetic induction, 
 at the point $\bx\in\RR^3$ at time $t\in\RR$. 
Let $\cI$ be a finite subset of the natural numbers $\NN$, of 
 cardinality $|\cI | = N$. 
We consider $N$ particles indexed by $\cI$. 
With Abraham, we assign to particle $k$ a
 rigid shape given by a nonnegative form function $f_k \in C_0^\infty(\RR^3)$,
 having  $SO(3)$ symmetry and satisfying $\int_{\RR^3}f_k(\bx){\rm d}^3x =1$.  
The total charge $q_{k}$ of particle $k$ is distributed 
 in space around the point $\by_k(t)\in \RR^3$
 by $\rho_k(\bx,t) = q_k f_k(\bx - \by_k(t))$, and 
 rigidly rotating with angular velocity $\bw_k(t)$. 
We also assign to the particle a `bare inertial mass' $m_{k}$ 
 (`material mass' in [\lorentzBOOKb]) and a `bare moment of inertia' $I_k$.
Moreover, $\nabla$ denotes the gradient operator with respect to $\bx$, 
 and a dot on top of a quantity will signify derivative with respect 
 to time, e.g. $\dot\by_k(t)$ is the linear velocity of particle $k$.

In semi-relativistic classical electron theory,
 the fields $\bE$ and $\bB$ satisfy the Maxwell-Lorentz equations
$$
\eqalignno{
{1\over c}\dif{\bB(\bx, t) }{t}  + \nabla\times\bE(\bx, t)  & = 0
&\eqlbl\maxwrotE\cr
 - {1\over c}\dif{\bE(\bx, t) }{t} 
+ \nabla\times\bB(\bx, t)  &= 4\pi {1\over c}\bj (\bx,t) 
&\eqlbl\maxwrotB\cr
\nabla\cdot \bB(\bx, t)  &= 0
&\eqlbl\maxwdivB \cr
\nabla\cdot\bE(\bx, t)  & = 4 \pi \rho (\bx, t) 
&\eqlbl\maxwdivE\cr}
$$
 where charge and current densities, $\rho(\bx, t)$ and $\bj(\bx,t)$, 
 are given by the Abraham-Lorentz expressions
$$
\eqalignno{
\rho (\bx, t) & = 
\sum_{k\in\cI} q_{k} f_{k}(\bx -\by_k(t))\, ,
&\eqlbl\ALcharge\cr
\bj (\bx,t) & = 
\sum_{k\in\cI} q_{k} f_{k}(\bx -\by_k(t))\bv_k(\bx,t) \, ,
&\eqlbl\ALcurrent\cr}
$$
 with
$$
\bv_k(\bx,t) = \dot\by_k(t) + \bom_k(t)\times (\bx -\by_k(t))\, .
\eqno\eqlbl\vfeld
$$
 The dynamical variables of particle $k$, momentum $\bp_k(t)\in \RR^3$ 
 and spin $\bs_k(t)\in \RR^3$, satisfy Newton's, respectively Euler's 
 equation of motion, equipped with the Abraham-Lorentz expressions for 
 the total force and torque acting on particle $k$, 
$$
\dot\bp_k = q_{k}\int_{\RR^3}
\left[\bE(\bx,t) + {1\over c}\bv_k(\bx,t)\times\bB(\bx,t)\right]
f_{k}(\bx -\by_k(t)) \, d^3x\, ,
\eqno\eqlbl\ALeqT
$$
$$
\dot\bs_k = q_{k}\int_{\RR^3} (\bx - \by_k(t))\times
\left[\bE(\bx,t) + {1\over c}\bv_k(\bx,t)\times\bB(\bx,t)\right]
f_{k}(\bx -\by_k(t)) \, d^3x\, .
\eqno\eqlbl\ALeqR
$$
Here,
$$
\bs_k = I_{k}\bom_{k} 
\eqno\eqlbl\sDEF
$$
 is the classical particle spin associated with the bare moment of 
 inertia, and   
$$
\bp_k = \cases{
\hskip 0.8truecm m_{k} \dot\by_{k}&{\rm (Newtonian)}\cr 
\displaystyle
{m_{k}\dot\by_{k}\over \sqrt{1- |\dot\by_{k}|^2/c^2}}&{\rm (Einsteinian)}\cr}
\eqno\eqlbl\pDEF
$$
 is the particle momentum associated with the bare mass.
Defining the  translational kinetic energy associated with the bare mass,
$$
T_k(\bp_k) = \cases{ \hskip 1.2truecm 
\displaystyle {1\over 2} {|\bp_{k}|^2 \over m_k} &{\rm (Newtonian)}\cr 
\displaystyle
{m_kc^2 \sqrt{1 + {|\bp_{k}|^2\over m_k^2c^2}}}&{\rm (Einsteinian)}$,$\cr}  
\eqno\eqlbl\TkDEF
$$
 we notice that velocity $\dot\by_k$ and momentum $\bp_k$ are, in either case, 
 related by
$$
\dot\by_k = {\partial T_k\over \partial \bp_k}\, .
\eqno\eqlbl\vkTk
$$
Both Newtonian [\abrahamBOOK,\lorentzBOOKb] and Einsteinian 
[\lorentzBOOKb,\bauerduerr,\komechspohn,\kunzespohnA,\spohn,\komechspohnkunze,\komechkunzespohn,\kunzespohnB]
 momenta 
 have been used in semi-relativistic variants of classical electron theory.
We therefore discuss both cases of (\pDEF), 
 but only the nonrelativistic Euler form (\sDEF) for spin.

Naturally, we want to treat these equations as a Cauchy problem, with
 initial data posed at time $t =t_0$. 
For the mechanical variables of the particles, the data are $\by_k(t_0)$, 
 $\dot\by_k(t_0)$, and $\bom_k(t_0)$; and for the fields, 
 $\bB(\bx,t_0)$ satisfying (\maxwdivB), and $\bE(\bx,t_0)$ 
 satisfying (\maxwdivE) at $t=t_0$. 
Actually, one should also  think of (\maxwdivB) and (\maxwdivE) 
 rather as initial conditions, to be imposed only at $t=t_0$, 
 on the initial data $\bB(\bx,t_0)$ and $\bE(\bx,t_0)$, for   
 the above set of equations is slightly redundant.  
In fact, (\maxwdivB) and (\maxwdivE) are automatically satisfied 
 for all $t$ if they are satisfied at $t=t_0$. 
For (\maxwdivB) this is seen by taking the divergence of (\maxwrotE). 
For (\maxwdivE) this is seen by taking the divergence of (\maxwrotB) 
 and the time-derivative of (\maxwdivE), then using the continuity 
 equation for the charge, which is proven below to hold as consequence 
 of (\ALcharge), (\ALcurrent), and (\vfeld) alone.  

Finally, a few remarks are in order regarding the bare inertias, 
 $m_k$ and $I_k$. 
By following up on Thomson's discovery [\jjthomsonA] that the 
 electromagnetic field of a particle contributes to its inertia, 
 Abraham [\abrahamBOOK] in particular, but also Lorentz
 [\lorentzBOOKb,\lorentzACAD],  suggested that inertia is 
 entirely due to electromagnetic effects, 
 and consequently proposed to set $m_k=0=I_k$ in (\pDEF) and (\sDEF). 
However, setting $m_k=0$ and/or $I_k=0$ is in
 serious conflict with the mathematical structure of a Cauchy problem, 
 see [\appelkiessling] for more on this.
In another vein, ever since Dirac's work [\dirac] there has been quite 
 some interest in letting $m_k\to -\infty$ associated with the 
 mass-renormalized point particle limit 
 $f_k(\bx-\by_k) \to \delta(\bx-\by_k)$, 
 see [\barutBOOKa,\rohrlichBOOK,\spohn].
However, stability problems emerge when $m_k<0$ and/or $I_k<0$, 
 cf. [\bauerduerr] for $m_k<0$ when only translational motions are 
 considered.
All these problems together suggest that one should choose the bare 
 inertias strictly positive, i.e. $m_k >0$ and $I_k>0$. 
Formally though, as can be seen upon inspection of our proofs below,
 the conservation laws hold for all regular solutions of 
 (\maxwrotE)--(\pDEF), with any values of $m_k\in\RR$ and $I_k\in \RR$. 

 \vfill\eject

\bigskip
\noindent{\bf III. CONSERVATION LAWS}
\chno=3\equno=0
\medskip

We assume that the initial conditions correspond to finite
 charge, total energy, linear and angular momentum. 
Then, because of the finite propagation speed for the electromagnetic 
 fields and the non-singular shape function, it is
 reasonable to expect (but not proven here) that the particle 
 speeds remain bounded and the motions and fields regular, so 
 that all  surface integrals over the fields at infinity 
 vanish at all times.
We now prove that, as a consequence of these hypotheses, the 
 traditional expressions of total charge, total energy, 
 total linear and total angular momentum, are conserved 
 quantities for	the dynamical equations (\maxwrotE)-(\pDEF).

\bigskip
\noindent{\bf IIIa. Charge conservation}
\medskip

{\it The total charge
$$
Q = \int_{\RR^3} \rho(\bx,t)\, d^3x  
\eqno\eqlbl\totQ
$$
is conserved. }

We need to show that $\dot Q =0$. For this it suffices to prove
 that the continutity equation 
$$
\dif{\rho(\bx,t)}{t} + \nabla\cdot \bj(\bx,t) = 0
\eqno\eqlbl\continEQ
$$
 is satisfied. 

We take the partial derivative of (\ALcharge) with respect to time, finding
$$
\dif{\rho}{t}(\bx,t) = - \sum_{k\in\cI} q_{k} 
\dot{\by}_k(t)\cdot\nabla f_{k}(\bx -\by_k(t))\, ,
\eqno\eqlbl\ddtALcharge
$$
 where we used that 
 $\dif{f_{k}(\bx -\by)}{\by} = -\dif{f_{k}(\bx -\by)}{\bx}$,
 with the identification 
 $\dif{f_{k}(\bx -\by)}{\bx} =\nabla{f_{k}(\bx -\by)} $. 
Next we take the divergence of (\ALcurrent) and obtain
$$
\nabla\cdot \bj (\bx,t)   =
\sum_{k\in\cI} q_{k} \Bigl(\bv_k(\bx,t)\cdot\nabla f_{k}(\bx -\by_k(t))
+ f_{k}(\bx -\by_k(t)) \nabla \cdot\bv_k(\bx,t)\Bigr) \, .
\eqno\eqlbl\divcurrent
$$
Noting that
$$
\bom_k(t)\times (\bx -\by_k(t)) = -
\nabla\times \left({1\over 2}\bom_k(t)\bigl|\bx - \by_k(t)\bigr|^2\right)\, ,
\eqno\eqlbl\diffsqr
$$
 it follows that
$$
\nabla \cdot\bv_k(\bx,t) = \nabla \cdot \Bigl(\dot\by_k(t) +
\bom_k(t)\times (\bx -\by_k(t))\Bigr) = 0\, ,
\eqno\eqlbl\divvnull
$$
 whence, 
$$
\nabla\cdot \bj (\bx,t)   =  \sum_{k\in\cI} q_{k} 
\dot{\by}_k(t)\cdot\nabla f_{k}\left(\bx -\by_k(t)\right)\, .
\eqno\eqlbl\divcurrentB
$$

In view of (\divcurrentB) and (\ddtALcharge), (\continEQ) holds. 
Thus, conservation of  charge (\totQ) is proved.


\bigskip
\noindent{\bf IIIb. Energy conservation}
\medskip

{\it The total energy 
$$
W = 
{1\over 8\pi} \int_{\RR^3} \bigl(|\bE|^2 + |\bB|^2\bigr)\, d^3x + 
\sum_{k\in\cI}\Bigl(T_k + {1\over 2I_k} |\bs_k|^2\Bigr)
\eqno\eqlbl\energy
$$
is conserved. }

We need to show that 
$$
\dot{W} = {1\over 4\pi} 
\int_{\RR^3} (\bE\cdot\dif{\bE}{t}
 + \bB\cdot\dif{\bB}{t})\, d^3x   
+ \sum_{k\in\cI}( \dot\by_k\cdot\dot\bp_k + 
\bom_k\cdot\dot\bs_k) = 0 \, .
\eqno\eqlbl\ddtenergy
$$

In the field integral in (\ddtenergy) we use (\maxwrotE) to 
 express $\dif{\bB}{t}$ in terms of $\nabla\times \bE$, 
 and (\maxwrotB) to express $\dif{\bE}{t}$ in terms of
 $\nabla\times \bB$ and $\bj$, then use the standard 
 identity (e.g. [\jacksonBOOK]) 
$$
\bB\cdot\nabla\times\bE - \bE\cdot\nabla\times\bB =
 \nabla\cdot (\bE\times\bB)\, ,
\eqno\eqlbl\rotEkreuzB
$$ 
 apply Gauss' theorem, notice that the surface integral 
 at infinity vanishes, and get
$$
\eqalignno{
 \int_{\RR^3} (\bE\cdot\dif{\bE}{t} + \bB\cdot\dif{\bB}{t})\, d^3x   
& 
= -\int_{\RR^3} ( c \nabla\cdot (\bE\times\bB) + 4\pi \bE\cdot\bj)\, d^3x
&\cr\cr&
= - 4\pi\! \int_{\RR^3} \bE\cdot\bj \,  d^3x   \, .
&\eqlbl\fieldWdotID\cr}
$$

As for the sum over particles, we insert the \rhs\ of (\ALeqT) for 
 $\dot\bp_k$, and the \rhs\ of (\ALeqR) for $\dot\bs_k$. 
We notice that (\ALeqT) contains a term $\perp\dot\by_k$, which vanishes 
 under the dot product with $\dot\by_k$ in (\ddtenergy). 
Similarly, inserting $\bX = \bx - \by_k$,\ $\bY =\bom_k$ and $\bZ=\bB$
 in the vector identity 
$\bX \times \Bigl((\bX\times\bY )\times\bZ\Bigr) = 
 (\bX\times \bY)(\bX \cdot\bZ)$,
 we see that (\ALeqR) contains a term $\perp\bom_k$, 
 which vanishes under the dot product with $\bom_k$ in (\ddtenergy). 
This, and a little vector algebra, gives us
$$
\eqalignno{
& 
\sum_{k\in\cI} (\dot\by_k\cdot\dot\bp_k +  \bom_k\cdot\dot\bs_k)  
&\cr&
\qquad = \sum_{k\in\cI}q_{k}\int_{\RR^3} 
 \biggl( \dot\by_k\cdot \left[\bE(\bx,t) 
+ {1\over c}\Bigl(\bom_k(t)\times (\bx -\by_k(t))\Bigr)\times\bB(\bx,t)\right]
&\cr& 
\qquad\qquad\qquad\quad
+ \bom_k\cdot \left\{(\bx - \by_k(t))\times \left[\bE(\bx,t) 
+ {1\over c}\dot\by_k(t) \times\bB(\bx,t)\right]\right\}  \biggr)  
f_{k}(\bx -\by_k(t)) 
\, d^3x 
&\cr&
\qquad = \sum_{k\in\cI}q_{k}\int_{\RR^3} f_{k}(\bx -\by_k(t)) 
 \Bigl( \dot\by_k + \bom_k(t)\times (\bx -\by_k(t))\Bigr)
\cdot \bE(\bx,t)  \, d^3x  
&\cr& 
\qquad = \int_{\RR^3} \sum_{k\in\cI}q_{k} f_{k}(\bx -\by_k(t)) 
 \bv_k(\bx,t) \cdot \bE(\bx,t)  \, d^3x \, .
&\eqlbl\WdotsumINTERM \cr}
$$
Recalling (\ALcharge) and (\ALcurrent), we finally get 
$$
\sum_{k\in\cI} (\dot\by_k\cdot\dot\bp_k +  \bom_k\cdot\dot\bs_k) 
 =  \int_{\RR^3} \bE\cdot\bj \, d^3x   \, .
\eqno\eqlbl\Wdotsum 
$$

With (\Wdotsum) and (\fieldWdotID) we see that the integral and sum 
 in (\ddtenergy) cancel in a manifest way, yielding $\dot W = 0$. 
Energy conservation is proved.

\bigskip
\noindent{\bf IIIc. Momentum conservation}
\medskip

{\it The total linear momentum
$$
\bP = 
{1\over 4\pi c} \int_{\RR^3} \bE\times\bB \, d^3x + 
\sum_{k\in\cI} \bp_k 
\eqno\eqlbl\linimpuls
$$
is conserved. }

We need to prove that
$$
\dot{\bP} = {1\over 4\pi c} \int_{\RR^3} (\bE\times\dif{\bB}{t}
 - \bB\times\dif{\bE}{t})\, d^3x   
+ \sum_{k\in\cI}  \dot\bp_k = {\bf 0}\, .
\eqno\eqlbl\ddtlinimpuls
$$

In the field integral in (\ddtlinimpuls) we use (\maxwrotE) to 
 express $\dif{\bB}{t}$ in terms of $\nabla\times \bE$
 and (\maxwrotB) to express $\dif{\bE}{t}$ in terms of
 $\nabla\times \bB$ and $\bj$, then integrate the standard 
 vanishing identity 
$$
{\bf 0} =  \bB\nabla\cdot\bB  + \bE \nabla\cdot\bE - 4\pi \rho \bE
\eqno\eqlbl\maxwIDa
$$
 over $\RR^3$, divide by $4\pi$, and add the result to our integral. 	
We next recall that
$$
\bE\nabla\cdot\bE + \bB\nabla\cdot\bB  - 
\bE\times\nabla\times\bE - \bB\times\nabla\times\bB =
4\pi \nabla\cdot \cM\, ,
\eqno\eqlbl\maxwMid
$$
 where 
$$
\cM = {1\over 4\pi} \left(\bE\otimes\bE + \bB\otimes\bB -
{1\over 2}\Bigl(|\bE|^2 + |\bB|^2\Bigr) {\cI}\right)
\eqno\eqlbl\maxwM
$$
 is Maxwell's symmetric stress tensor, with ${\cI}$ 
 the identity $3\times 3$ tensor.
But
$$
\int_{\RR^3}\nabla\cdot \cM \, d^3x  = {\bf 0}\, ,
\eqno\eqlbl\maxwMdiv
$$
 since, by Gauss' theorem, the integral on the left 
 can be transformed into a surface integral at $\infty$, 
 where it vanishes, by our hypotheses.
Thus, we have
$$
{1\over 4\pi c} \int_{\RR^3} (\bE\times\dif{\bB}{t}
 - \bB\times\dif{\bE}{t})\, d^3x   =
- \int_{\RR^3} \Bigl( \rho\bE + {1\over c}\bj\times \bB\Bigr)\, d^3x
\, ,
\eqno\eqlbl\dotPintONE
$$
 with $\rho$ given by (\ALcharge), and $\bj$ by (\ALcurrent). 

As for the sum over particles, we insert the \rhs\ of (\ALeqT) for 
 $\dot\bp_k$, then exchange summation and integration, 
 recall (\ALcharge) and (\ALcurrent), and obtain, 
$$
\eqalignno{
\sum_{k\in\cI} \dot\bp_k 
& 
=
\int_{\RR^3} \sum_{k\in\cI} q_{k}f_{k}(\bx -\by_k(t)) 
\left[\bE(\bx,t) + {1\over c}\bv_k(\bx,t)\times\bB(\bx,t)\right]\,d^3x 
&\cr&
= \int_{\RR^3} \Bigl( \rho\bE + {1\over c}\bj\times \bB\Bigr)\, d^3x
\, .
&\eqlbl\summomentumID\cr}
$$

With (\dotPintONE) and (\summomentumID) inserted into 
 (\ddtlinimpuls), we obtain $\dot{\bP} = {\bf 0}$. 
Linear momentum conservation is proved.

\bigskip
\noindent{\bf IIId. Angular momentum conservation}
\medskip

{\it The total angular momentum
$$
\bJ  = 
{1\over 4\pi c} \int_{\RR^3} \bx\times (\bE\times\bB) \, d^3x + 
\sum_{k\in\cI}( \by_k\times\bp_k + \bs_k )
\eqno\eqlbl\drehimpuls
$$
is conserved. }

We need to show that 
$$
\dot{\bJ} = {1\over 4\pi c} \int_{\RR^3} \bx\times 
(\bE\times\dif{\bB}{t}  - \bB\times\dif{\bE}{t})\, d^3x   +
 \sum_{k\in\cI}( \by_k\times\dot\bp_k + \dot\bs_k ) ={\bf 0}\, .
\eqno\eqlbl\ddtdrehimpuls
$$

Turning first to the field integral in (\ddtdrehimpuls), we use 
 (\maxwrotE) to express $\dif{\bB}{t}$ in terms of $\nabla\times \bE$
 and (\maxwrotB) to express $\dif{\bE}{t}$ in terms of
 $\nabla\times \bB$ and $\bj$. Next we take the cross
 product of (\maxwIDa) with $\bx$, obtaining the vanishing identity 
$$
0 = \bx\times ( \bB\nabla\cdot\bB  + \bE\nabla\cdot\bE - 4\pi \rho\bE)
\, .
\eqno\eqlbl\maxwIDb
$$
 We integrate (\maxwIDb) over $\RR^3$, divide by $4\pi$, 
 and add the result to our integral in (\ddtdrehimpuls). 
With the help of (\maxwMid), this gives us 
$$
{1\over 4\pi c} \int_{\RR^3} \bx\times 
(\bE\times\dif{\bB}{t}  - \bB\times\dif{\bE}{t})\, d^3x   =
 \int_{\RR^3} \bx\times\Bigl(\nabla\cdot\cM -  \rho\bE
- {1\over c}\bj\times \bB\Bigr)\, d^3x  \, ,
\eqno\eqlbl\maxwIDc
$$
 with $\rho$ given by (\ALcharge) and $\bj$ by (\ALcurrent). 
Finally, recalling the identity [\jacksonBOOK]
$$
 \int_{\RR^3} \bx\times \nabla\cdot\cM \, d^3x  
= 
\lim_{R\to\infty}\oint_{\SS^2_R} 
(\cM \cdot \bx)\times\bn\, d\sigma = {\bf 0}\, ,
\eqno\eqlbl\MidGauss
$$
 (where we also used that $\nabla\times\bx ={\bf 0}$), we finally have
$$
{1\over 4\pi c} \int_{\RR^3} \bx\times 
\bigl(\bE\times\dif{\bB}{t}  - \bB\times\dif{\bE}{t}\bigr)\, d^3x   =
- \int_{\RR^3} \bx\times\Bigl(\rho\bE
+ {1\over c}\bj\times \bB\Bigr)\, d^3x  \, .
\eqno\eqlbl\drehimpulsint
$$

Coming to the sum over particles, we insert the \rhs\ of (\ALeqT) 
 for $\dot\bp_k$, the \rhs\ of (\ALeqR) for 
 $\dot\bs_k$, notice some obvious cancelations, and obtain
$$
\eqalignno{
\sum_{k\in\cI}
&( \by_k\times\dot\bp_k + \dot\bs_k ) 
&\cr&
= \sum_{k\in\cI}q_{k} \Bigl( \by_k\times \int_{\RR^3}
\left[\bE(\bx,t) + {1\over c}\bv_k(\bx,t)\times\bB(\bx,t)\right]
f_{k}(\bx -\by_k(t)) \, d^3x\ 
&\cr&
\qquad  + \int_{\RR^3} (\bx - \by_k(t))\times
\left[\bE(\bx,t) + {1\over c}\bv_k(\bx,t)\times\bB(\bx,t)\right]
f_{k}(\bx -\by_k(t)) \, d^3x
\Bigr) 
&\cr&
 =\, \sum_{k\in\cI}q_{k} 
  \int_{\RR^3} \bx \times
\left[\bE(\bx,t) + {1\over c}\bv_k(\bx,t)\times\bB(\bx,t)\right]
f_{k}(\bx -\by_k(t)) \, d^3x\, .
&\eqlbl\impulssumcancel\cr}
$$
In the last expression we can exchange summation and integration. 
Recalling (\ALcharge) and (\ALcurrent), we find 
$$
 \sum_{k\in\cI}
( \by_k\times\dot\bp_k + \dot\bs_k )
= \int_{\RR^3} \bx\times\Bigl(\rho\bE
+ {1\over c}\bj\times \bB\Bigr)\, d^3x  \, .
\eqno\eqlbl\drehimpulssum
$$

With (\drehimpulssum) and (\drehimpulsint) inserted in
 (\ddtdrehimpuls), we have $\dot{\bJ} =0$. 
Conservation of total angular momentum is proved. 

\bigskip
\noindent{\bf IV. NON-CONSERVATION OF ANGULAR MOMENTUM 
WHEN $\bom\equiv {\bf 0}$}
\chno=4\equno=0
\medskip

It is now easily seen that, upon setting $\bom_k \equiv {\bf 0}$ everywhere
 in the equations of motion, the traditional 
 expressions for charge, energy and linear momentum are 
 still conserved, but the one for angular momentum is not. 
Indeed, if in the equations of motion we set $\bom_k \equiv {\bf 0}$ 
 for all $k$, and then follow through the computations of section
 III step by step, with $\bom_k\equiv {\bf 0}$ in place
 everywhere, we easily verify that the conclusions of 
 subsections III.a, III.b, and III.c still hold.	
However, if we go through the steps of subsection III.d, 
 with $\bom_k\equiv {\bf 0}$ in place everywhere, we obtain 
$$
\dot\bJ = -  \sum_{k\in\cI} q_k
\int_{\RR^3} (\bx - \by_k(t))\times
\left[\bE(\bx,t) + {1\over c}\dot\by_k(t)\times\bB(\bx,t)\right]
f_{k}(\bx -\by_k(t)) \, d^3x\, .
\eqno\eqlbl\EXTRAtorque
$$
The right side in (\EXTRAtorque) is, in general, an uncompensated 
 sum of torques. 
Hence, except for some special highly symmetric situations, there will
 be a non-vanishing rate of change of total angular momentum.

\bigskip
\noindent{\bf V. CLOSING REMARK}
\chno=5\equno=0
\medskip

Our observation could have been made at the beginning of the $20^{th}$
 century, by Abraham, Lorentz or Poincar\'e, but apparently it wasn't.  
So it was left to Uhlenbeck and Goudsmit [\uhlenbeckgoudsmit] to
 re-invent particle spin for the interpretation of spectral data.
It is amusing to contemplate that the story of spin [\tomonagaBOOK] 
 could have been a different one.   
\medskip
\noindent
{\bf ACKNOWLEDGEMENT}. I thank Eugene Speer for reading the
	manuscript. Work supported by NSF GRANT DMS-9623220. 

\vfill\eject

\biblio

\medskip
\noindent
{Submitted: April 29, 1999; Accepted: May 18, 1999.}

\bye